\documentclass[manuscript]{aastex}
\usepackage{epsfig}

\slugcomment{Submitted to the Astronomical Journal}

\shorttitle{Orbital Solutions for Two Young, Low-Mass Spectroscopic Binaries in Ophiuchus}
\shortauthors{Rosero et al.}

\begin{document}

\title{Orbital Solutions for Two Young, Low-Mass Spectroscopic Binaries in Ophiuchus}

\author{V. Rosero\altaffilmark{1}}
\affil{Lowell Observatory, 1400 West Mars Hill Road,
    Flagstaff, AZ 86001, USA}
\email{viviana@lowell.edu}

\author{L. Prato}
\affil{Lowell Observatory, 1400 West Mars Hill Road,
 Flagstaff, AZ 86001, USA}
\email{lprato@lowell.edu}

\author{L. H. Wasserman}
\affil{Lowell Observatory, 1400 West Mars Hill Road,
 Flagstaff, AZ 86001, USA}
\email{lhw@lowell.edu}

\and

\author{B. Rodgers}
\affil{Gemini Observatory, Gemini South, 
AURA/Chile, P.O. Box 26732,
Tucson, AZ 85726, USA}
\email{brodgers@gemini.edu }

\altaffiltext{1}{Current address: Physics Department, 333 Workman Center,
New Mexico Institute of Mining and Technology, 801 Leroy Place, Socorro, NM 87801}

\begin{abstract}
We report the orbital parameters for ROXR1 14 and RX J1622.7-2325Nw, two young, low-mass, and double-lined spectroscopic binaries recently discovered in the Ophiuchus star forming region. Accurate orbital solutions were determined from over a dozen high-resolution spectra  taken with the Keck II and Gemini South telescopes. These objects are T Tauri stars with mass ratios close to unity and periods of $\sim$5 and $\sim$3 days, respectively. In particular, RX J1622.7-2325Nw shows a non-circularized orbit with an eccentricity of 0.30, higher than any other short-period pre-main sequence spectroscopic binary known to date. We speculate that orbit of RX J1622.7-2325Nw has not yet circularized because of the perturbing action of a $\sim1''$ companion, itself a close visual pair. A comparison of known young spectroscopic binaries and main sequence (MS) spectroscopic binaries in the eccentricity-period plane shows an indistinguishable distribution of the two populations, implying that orbital circularization occurs in the first 1 Myr of a star's lifetime. With the results presented in this paper we are increasing by $\sim4\%$  the small sample of PMS spectroscopic binary stars with known orbital elements. 

\end{abstract}


\section{INTRODUCTION}
T Tauri stars (TTS) are very young (1 to 10 Myrs), low-mass pre-main sequence (PMS) stars. Low-mass stars are the most abundant stellar objects in the Galaxy (e.g., Bean et al. 2006), therefore it is important to understand their formation and evolution.  Spectroscopic binaries (SB) among young stars are important for the information that they potentially provide for the study of stellar parameters, such as the possibility to make dynamical measurements of mass ratios and eventually the masses of the  components (Prato et al. 2002a).   The mass and chemical composition of a star at birth determines its entire evolution. Accurate determination of PMS stellar masses is extremely important for refining the current young star theoretical models, which are not well calibrated by empirical data (Hillenbrand \& White 2004). Therefore, the study of low-mass T Tauri SBs is crucial for understanding star formation.  \\

Despite  their importance, only a few young, low-mass SBs are well-studied. Melo et al. (2001) and Prato  et al. (2002b) presented compilations of $\sim$49 systems with
known orbital elements. This sample has increased by fewer than 10 objects in the last decade. The small number of known PMS SBs is in part
attributable to their location in obscured and distant star
forming regions: it can be an arduous  task to measure precise radial velocities for these
low-mass stars with the traditional approach of spectroscopy in visible light.  Furthermore, low-mass T Tauri stars generally
emit most of their energy at wavelengths $\geq 1\, \mu m$ and thus the most efficient way to observe them is with large-aperture telescopes in the infrared (IR, Prato 2007).  This has only become possible in the last decade.\\

ROXR1 14 and RX J1622.7-2325Nw are PMS double-lined spectroscopic binary (SB2)  stars, located in Ophiuchus, originally identified in a survey of T Tauri M stars (Prato 2007).  Results presented here are from the analysis of over a dozen spectra for each object obtained with high-resolution IR spectrographs. In this paper we report on the measurement of radial velocities using two-dimensional cross-correlation in order to calculate the  periods and orbital parameters of these PMS binary systems. Therefore, we are increasing by two objects the sample of PMS spectroscopic binaries with well-determined parameters.\\

The layout of the paper is the following. In \S \ref{observations}, we present the observations and data reduction procedure. In  \S \ref{rv}, we provide the radial velocity calculations. Results for both objects are presented in \S \ref{results}. A discussion appears in \S \ref{discussions} and a summary in \S \ref{sum}.

\section{OBSERVATIONS AND DATA REDUCTION}\label{observations}
Spectra of the target stars were taken with the NIRSPEC and Phoenix spectrometers on Keck II and Gemini South, respectively. The spectral range of Phoenix is smaller than that of NIRSPEC, however, its resolution is $\sim 2$ times greater. The two instruments provided similar precision for the radial velocity measurements presented here because these properties, in conjunction with the higher signal to noise ratio of the NIRSPEC data, tended to cancel out.  Basic target star properties are listed in Table \ref{target_prop} and the UT date of observation and instrument used appear in the last column of Tables \ref{roxdata} and \ref{rxjdata}.  Adaptive optics (AO) images were also taken of both target systems with the NIRC2 camera on the Keck II telescope.

\subsection{NIRSPEC}

\subsubsection{Observations}
From UT 2002 March to 2009 July observations were taken in different epochs  using NIRSPEC, the facility high-resolution, near-IR spectrograph located at the Keck II telescope on Mauna Kea (McLean et al. 1998, 2000).  The spectroscopic detector in this cryogenic, cross-dispersed instrument is an Aladdin InSb 1024 $\times$ 1024 array. We observed with a $0.288$'' slit, yielding resolution $R=30,000$. In our H-band setting, $\lambda = 1.555\, \mu m$ appears in the center of the middle order (order 49).  This order is advantageous because it is effectively devoid of atmospheric absorption lines, allowing for the analysis of data without the need for division of the spectra by a telluric standard star spectrum, thus optimizing the signal to noise ratio.  Also, there are numerous, strong OH night sky emission lines spaced across this order (Rousselot et al. 2000), providing an excellent inherent wavelength calibration. We used these lines to determine the dispersion solution. Of the 10 OH lines shown in Rousselot et al. (2000; Figure 16) that fall in order 49, two are close blends that NIRSPEC does not resolve. A well-resolved line at $1.5631443\, \mu m$ produces a large residual, so we did not use this one either.  We took  at least four spectra of every target each night with the telescope nodded between two positions on the slit in an A-B-B-A nod sequence in order to compute the background sky subtraction. Typical exposure times were 5 minutes and the resulting signal to noise ratio was usually $>200$.\\

\subsubsection{Data Reduction}\label{procedure}

Data were reduced using the REDSPEC package\footnote{See http://www2.keck.hawaii.edu/inst/nirspec/redspec.html}. Ten dark frames were median filtered in order to get a master dark. The same procedure was used to get a master flat; the difference, flat-dark, was used for subsequent analysis. High-resolution NIRSPEC H-band observations cover 9 orders, however, only the order centered on $1.555\, \mu m$ was isolated for the reduction and analysis. Additive terms such as  sky background were eliminated with a pair subtraction of our observing pattern, i.e. A-B. This difference was then divided by the dark subtracted flat. REDSPEC then spatially and spectrally rectified and calibrated the resulting difference. Both positive and negative spectra were  extracted. These sometimes contain residual OH night sky lines; by subtracting the negative from the positive extracted spectra and dividing by 2, these spectra were averaged and the OH residuals eliminated.  Lastly, the resulting spectrum was normalized and divided by a 2nd order polynomial fit to the continuum to flatten the final product. All spectra were corrected for heliocentric motion using the SkyCalc software package (Thorstensen 2001). \\

\subsection{Phoenix}

\subsubsection{Observations}
From UT 2008 March to 2008 June, high-resolution IR spectra were obtained on the 8.1 m Gemini South Telescope with the cryogenic Phoenix echelle spectrograph  in queue mode (GS-2008A-Q-7). With the $0.34$'' slit, the resolution was  $R=50,000$ (Hinkle et al. 2003). The Phoenix detector is an  Aladdin 512 $\times$ 1024 InSb array. Utilizing queue observing allowed us
to obtain several observations over just a few months. We observed in the H-band from $1.552-1.559\,  \mu m$, a subset of the wavelength range covered by NIRSPEC. Because Phoenix is not a cross dispersed instrument it provides only a single order. Four spectra for every target were taken with an A-B-B-A nod sequence in order to compute the background sky subtraction. Typical exposure times were 4 minutes and the resulting signal to noise ratio was usually $>130$. \\

\subsubsection{Data Reduction}
\begin{figure}[!h]
\centering
\includegraphics[scale=.60]{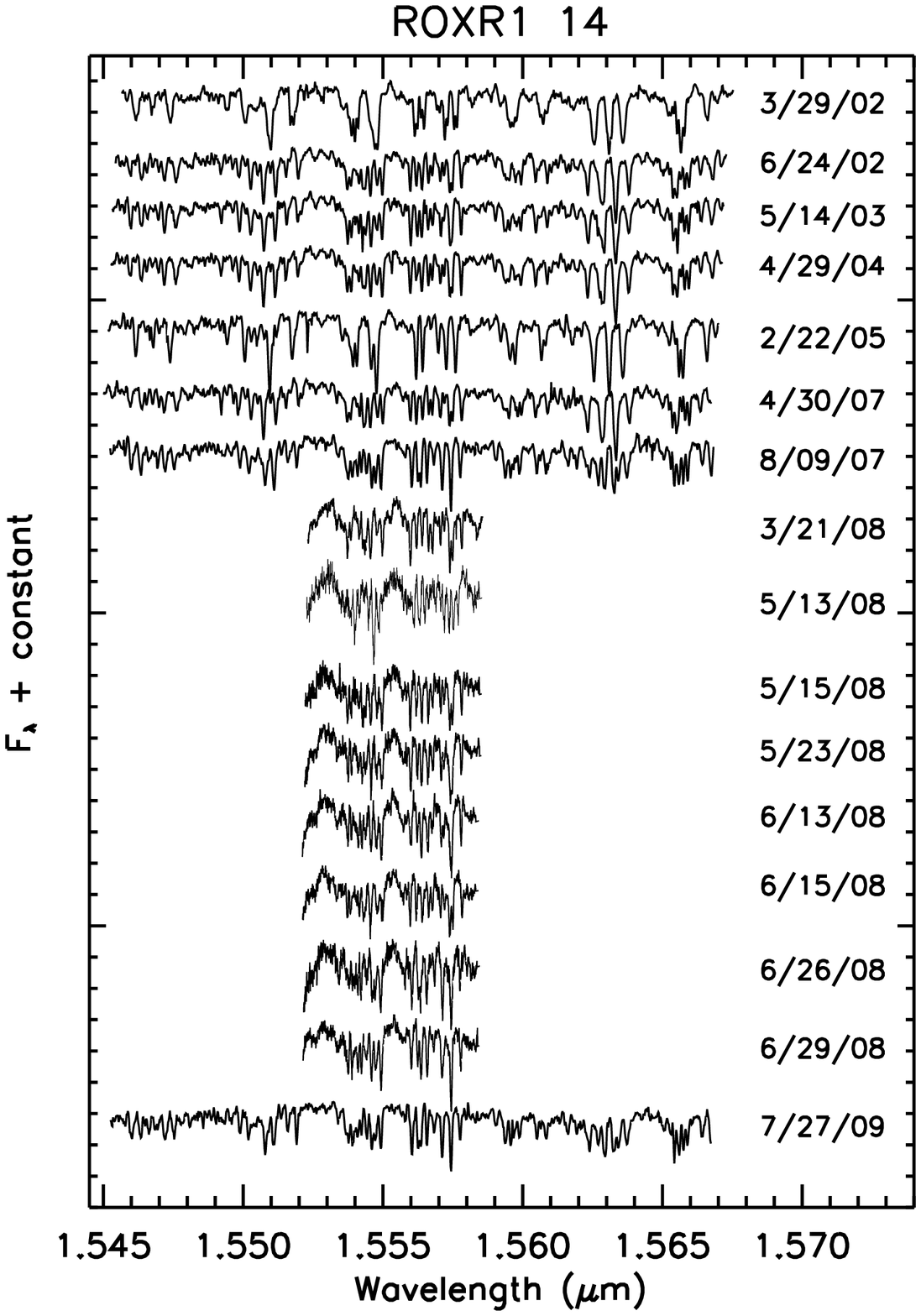}
\caption{NIRSPEC and Phoenix spectra of the SB2 ROXR1 14. All spectra have been corrected for heliocentric motion, normalized to a continuum level of unity, and separated in the y-axis by an additive constant. In this type of young PMS star, absorption lines are generally due to Fe, Ti, Ni, Si, OH and H$_{2}$O. }\label{rox_spectra}
\end{figure}

\begin{figure}[!h]
\centering
\includegraphics[scale=.60]{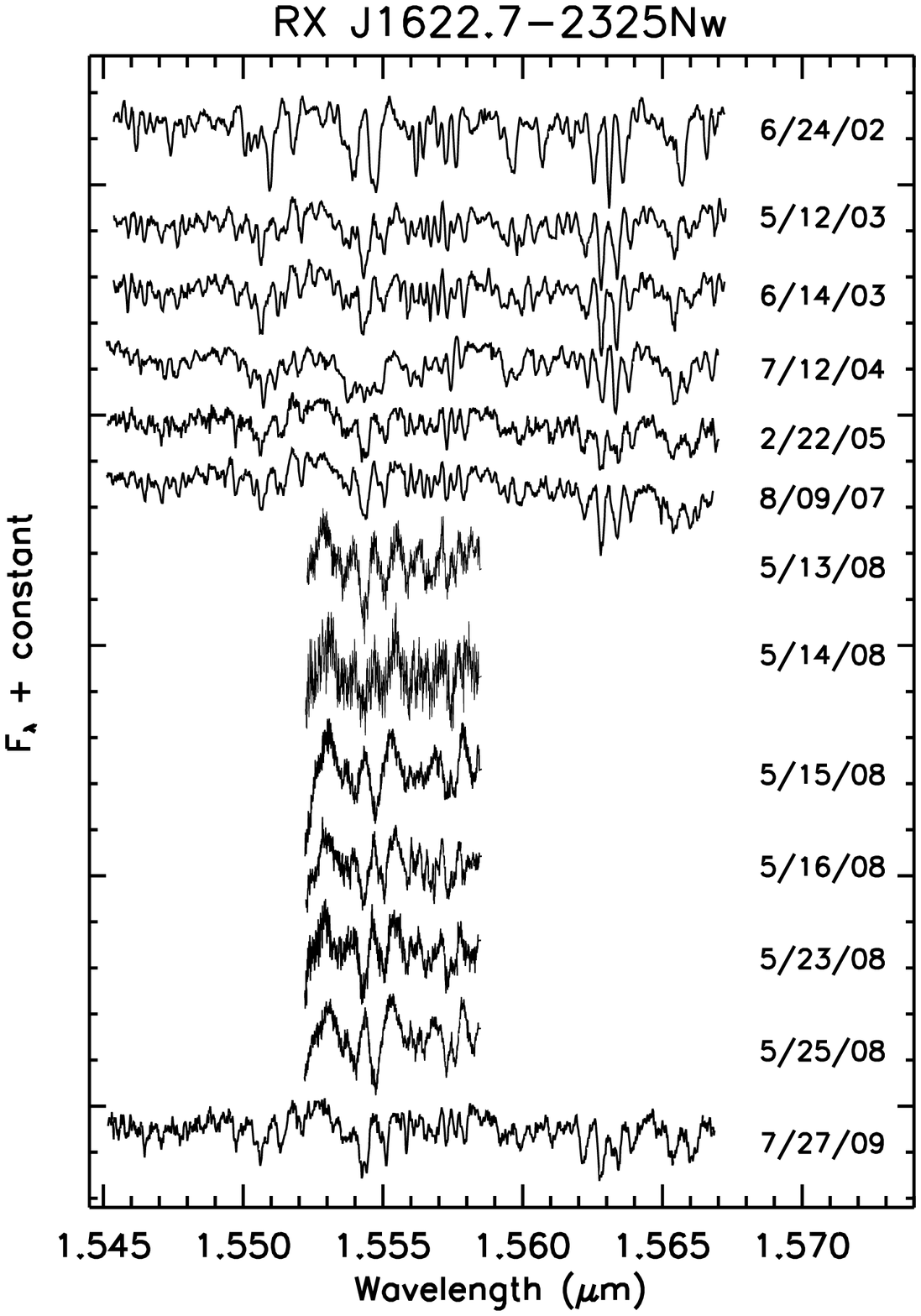}
\caption{NIRSPEC and Phoenix spectra of the SB2 RX J1622.7-2325Nw. All spectra have been corrected for heliocentric motion, normalized to a continuum level of unity, and separated in the y-axis by an additive constant.  In this type of young PMS star, absorption lines are generally due to Fe, Ti, Ni, Si, OH and H$_{2}$O. }\label{rxj_spectra}
\end{figure}

The same reduction algorithms used on the NIRSPEC data and the procedure described in \ref{procedure}  were applied to the Phoenix data. In order to align the data from Phoenix with the input specification of REDSPEC it was necessary to rotate the raw files by  90$^{ \circ}$. Reduced NIRSPEC and Phoenix spectra are shown in Figures \ref{rox_spectra} and \ref{rxj_spectra}.\\

\subsection{NIRC2}

On UT 2005 February 25 AO images of both targets were obtained on the Keck II telescope with the NIRC2 facility near-IR camera
and the K$_{cont}$ filter.
The NIRC2 detector is a 1024 $\times$ 1024 Aladdin-3 InSb array (Wizinowich et al. 2000); we used the narrow-field camera, with a plate scale of $\sim$0.01$''$. For each target we obtained one set of dithered images using a five-point pattern, integrating for
1 second at each position, with an offset 2$''$.  Pairs of images were subtracted from each other to remove sky background and visually examined for companion objects.

\section{Radial Velocity Calculations}\label{rv}

Radial velocities  were determined, following Zucker \& Mazeh (1994), from a two dimensional cross-correlation algorithm developed at Lowell Observatory. The algorithm calculates the correlation of the target  spectroscopic binary spectrum against  combinations of two templates, thus identifying the radial velocities and approximate flux ratio, $\alpha$,  of the primary and  the secondary components ($\alpha$ is $\sim 0.90$ for both of the systems analyzed in this paper). The rotational velocity, ${\it v}$sin\emph i, was estimated by using a set of templates convolved with a nonlinear limb-darkened broadening kernel (Bender \& Simon 2008) to produce a range of ${\it v}$sin\emph i values. We selected the ${\it v}$sin\emph i of the template yielding the maximum correlation. Different spectral type templates were also tested; however, as radial velocities are relatively insensitive to spectral type differences of one or two subclasses (Simon \& Prato 2004) and because the components have mass ratios close to unity (see below), the spectral types appearing in Table \ref{target_prop} are those identified in Prato 2007.\\

The standard stars used as templates are well-known, main sequence (MS) stars with a spectral type similar to that of the target. Although this is not ideal, for example the surface gravities of young stars are generally lower than those of MS stars, it is nonetheless effective for cross-correlation because the line positions dominate the results (e.g., Schaefer et al. 2008). Parameters for the standard stars used as templates appear in Table \ref{standard_st}. The template GL 763 (M0) yielded the best correlation for both of the components of ROXR1 14  with ${\it v}$sin\emph i$=15\pm 3\, km\,s^{-1}$. In contrast, for  RX J1622.7-2325Nw, BS 8086 (K7) gave the best correlation for the primary component and GL 763 for the secondary, with  ${\it v}$sin\emph i$=25\pm 5\, km\,s^{-1}$ for both. The systematic uncertainty for the radial velocities of  the templates is $\pm 1.0$ $km\,s^{-1}$, adopted from Prato et al. (2002b) and Steffen et al. (2001).\\

The component mass ratio impacts the uncertainties of the measured radial velocities; measurements of the secondary star RV in very small mass ratio systems can be challenging. In addition, the shape of the  peak and therefore the uncertainties  in the correlation depend strongly on the  rotational velocities of the primary and secondary, i.e. ${\it v}$sin\emph i.  Following Kurtz et al. (1992), internal errors determined from the cross-correlation algorithm were calculated as a function of the FWHM of the correlation peak and the ratio of the correlation peak height to the amplitude of the antisymmetric noise. To determine the total uncertainty of each target RV measurement, we combined the internal errors in quadrature with the systematic uncertainty introduced by the templates, $1\,km\,s^{-1}$.
Measured radial velocities of the primary, $v_{1}$, and the secondary, $v_{2}$, as well as their total uncertainties, $\sigma_{v_{1}}$ and $\sigma_{v_{2}}$, are reported in Tables \ref{roxdata} and \ref{rxjdata}.

\section{Results}\label{results}
The PMS double-lined spectroscopic binaries reported here were discovered by Prato (2007) during a survey of K7 and early M stars in search of young spectroscopic binaries. The survey targets were drawn from an X-ray selected  sample (Casanova et al. 1995;  Mart\'in et al. 1998) located in the $\rho$ Oph (L1688) molecular cloud. Both targets were classified as WTTS by Mart\'in et al. (1998)  using their spectral types and equivalent widths of H$\alpha$ and LiI $\lambda 6708\, \AA$. These properties are listed in Table \ref{target_prop}. Magnitudes were obtained from 2MASS and equivalent widths are from Mart\'in et al. (1998).  \\

\begin{figure}[!h]
\centering
\includegraphics[scale=.60]{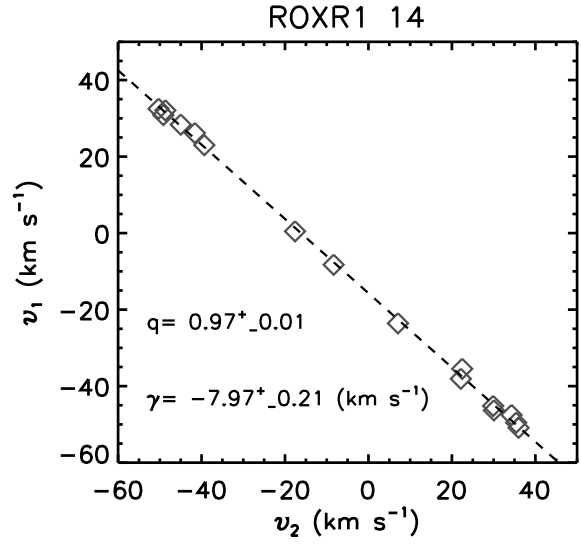}
\caption{Linear fit to the primary vs. secondary radial velocities for ROXR1 14. Uncertainties in the RV are not shown because they are smaller than the symbol size.}\label{rox_wilson}
\end{figure}

\begin{figure}[!h]
\centering
\includegraphics[scale=.60]{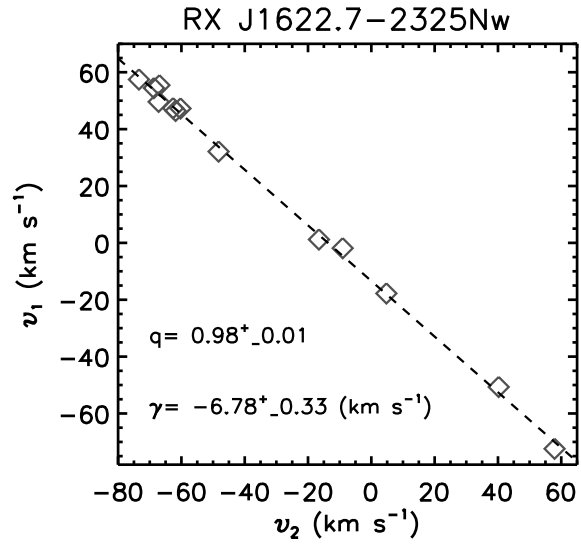}
\caption{Linear fit to the primary vs. secondary radial velocities for RX J1622.7-2325Nw. Uncertainties in the RV are not shown because they are smaller than the symbol size.}\label{rxj_wilson}
\end{figure}

For a preliminary result and consistency check, we derived mass ratios and center-of-mass velocities for the targets using the method of  Wilson (1941). For the different epochs, we plot the radial velocities of the primary versus the secondary (Figures \ref{rox_wilson} and \ref{rxj_wilson}). From the linear fit, the negative of the slope gives the mass ratio, q. The center-of-mass velocity $\gamma$ is given by the y-intercept divided by (1+q). \\

Initial guesses for the orbital solution of the SB2s were found using an amoeba search routine from Press et al (1992). Actual orbital solutions for each target  were determined with the Levenberg-Marquardt method  (Press et al. 1992) through  standard non-linear least-squares techniques. In some cases, velocities for the primary and the secondary were switched, since  the brightness of the component stars in the analyzed  systems are nearly the same. In those situations, orbital fits were calculated again after reassigning velocities  for the primary and the secondary.  The orbital parameters obtained from the orbital fit solution are period, P, the projected  semi-mayor axes of the components, $a_{1}sin$\emph i and $a_{2}sin$\emph i, eccentricity, {\it  e},  periastron angle,   $\omega$, time of periastron passage, T. Table \ref{parameters} shows the orbital parameter values and the derived parameters such as the semi-amplitude of both components, $K_{1}$ and $K_{2}$ and the mass ratio, $q=K_{1}/K_{2}$. Uncertainties calculated by the Levenberg-Marquardt method are the formal errors from the non-linear least-squares fitting.\\

\subsection{ROXR1 14}
\begin{figure}[!h]
\centering
\includegraphics[scale=.60]{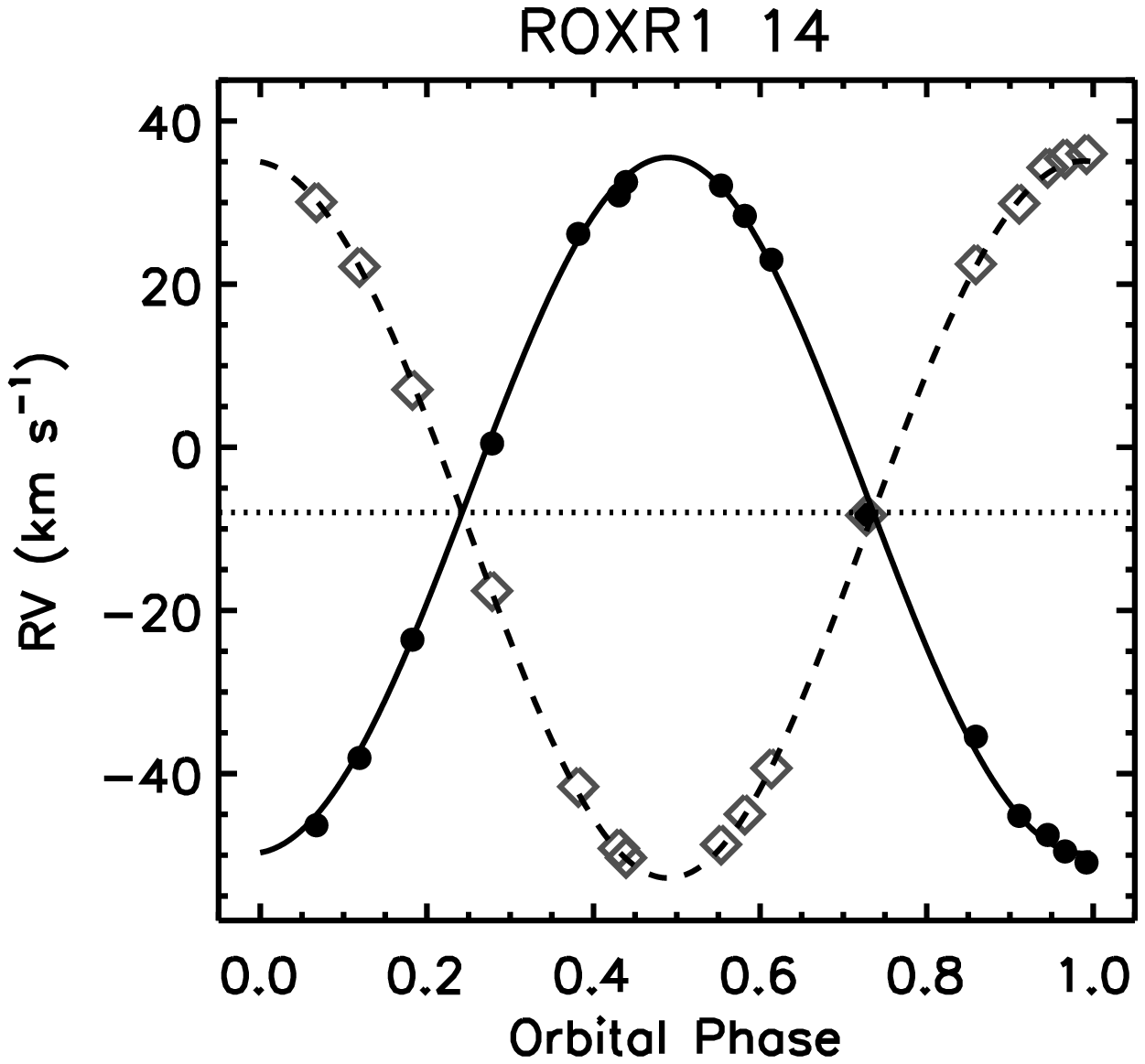}
\caption{Radial velocity as a function of phase for ROXR1 14. Measured radial velocities are represented as circles for the primary and as diamonds for the secondary. The dotted line shows  the center of mass velocity, $\gamma$. Uncertainties in the RV are not shown because they are smaller than the symbol size.}\label{rox_curve}
\end{figure}

ROXR1 14 is a SB2 with a period $P=5.72$ days  and  an eccentricity $e=0.020\pm 0.007$. The radial velocity curve as a function of the phase is shown in Figure \ref{rox_curve}.  Column 1 in Table \ref{parameters} shows the complete orbital parameter values for this object. The mass ratio derived from the semi-amplitude of the components' velocities is $0.97\pm 0.01$ and the center-of-mass velocity is  $\gamma=-7.98 \pm 0.18 $ $km\,s^{-1}$. These values are consistent with the ones derived from Figure \ref{rox_wilson} to within $\ll$1$\sigma$.  AO images of this system show no obvious evidence for a stellar companion.

\subsection{RX J1622.7-2325Nw}
\begin{figure}[!h]
\centering
\includegraphics[scale=.60]{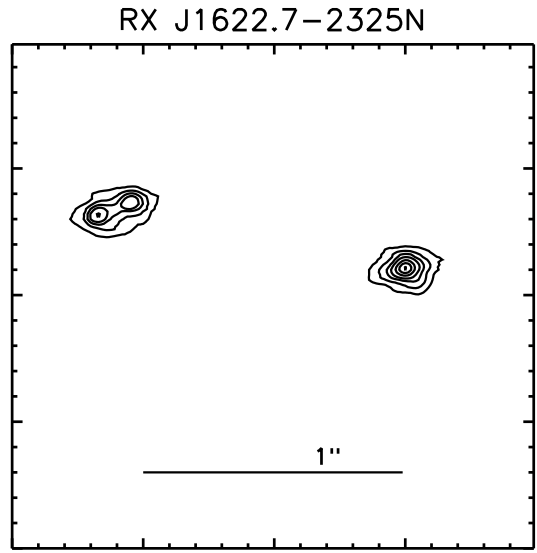}
\caption{K-band AO image of the RX J1622.7-2325N system taken in February, 2005 with the NIRC2 camera on Keck II. North is up and east is to the left.  The western component is the PMS SB2 and the eastern component is the visual binary (0.1'').}\label{AO_contour}
\end{figure}

RX J1622.7-2325Nw is the SB2 component of the hierarchical quadruple system identified  by Prato (2007) . The western component (Nw) is separated by $\sim$1$''$ from  the eastern one (Ne), which is itself a visual binary ($\sim$0.1$''$; see Figure \ref{AO_contour}).  The radial velocity curve as a function of the phase is shown in Figure \ref{rxj_curve}. Orbital parameters are shown in the second column of Table \ref{parameters}.  From the orbital solution we found $q=0.98\pm 0.06$ and $\gamma=-6.75\pm 1.09$ $km\,s^{-1}$, consistent with the values found in the Wilson plot shown in Figure \ref{rxj_wilson}. With a $P=3.23$ day period and $e=0.30\pm 0.04$, RX J1622.7-2325Nw  is the shortest-period high-eccentricity PMS SB known to date. 

\section{DISCUSSION}\label{discussions}

As expected, our result shows that stars with high projected rotational velocities, ${\it v}$sin\emph i, as is the case  of RX J1622.7-2325Nw (${\it v}$sin\emph i$=25\pm 5\, km\,s^{-1}$), will have larger uncertainties in the measured radial velocities (Table \ref{rxjdata} columns 3 and 5) compared with the smaller errors in the radial velocities of more slowly rotating stars such as ROXR1 14 (${\it v}$sin\emph i$=15\pm 3\, km\,s^{-1}$; Table \ref{roxdata} columns 3 and 5). The reason for this is that the uncertainties in the radial velocities are related to the FWHM of the cross-correlation peak, which is wider for stars with larger ${\it v}\,sin\emph i$. \\

\begin{figure}[!h]
\centering
\includegraphics[scale=.60]{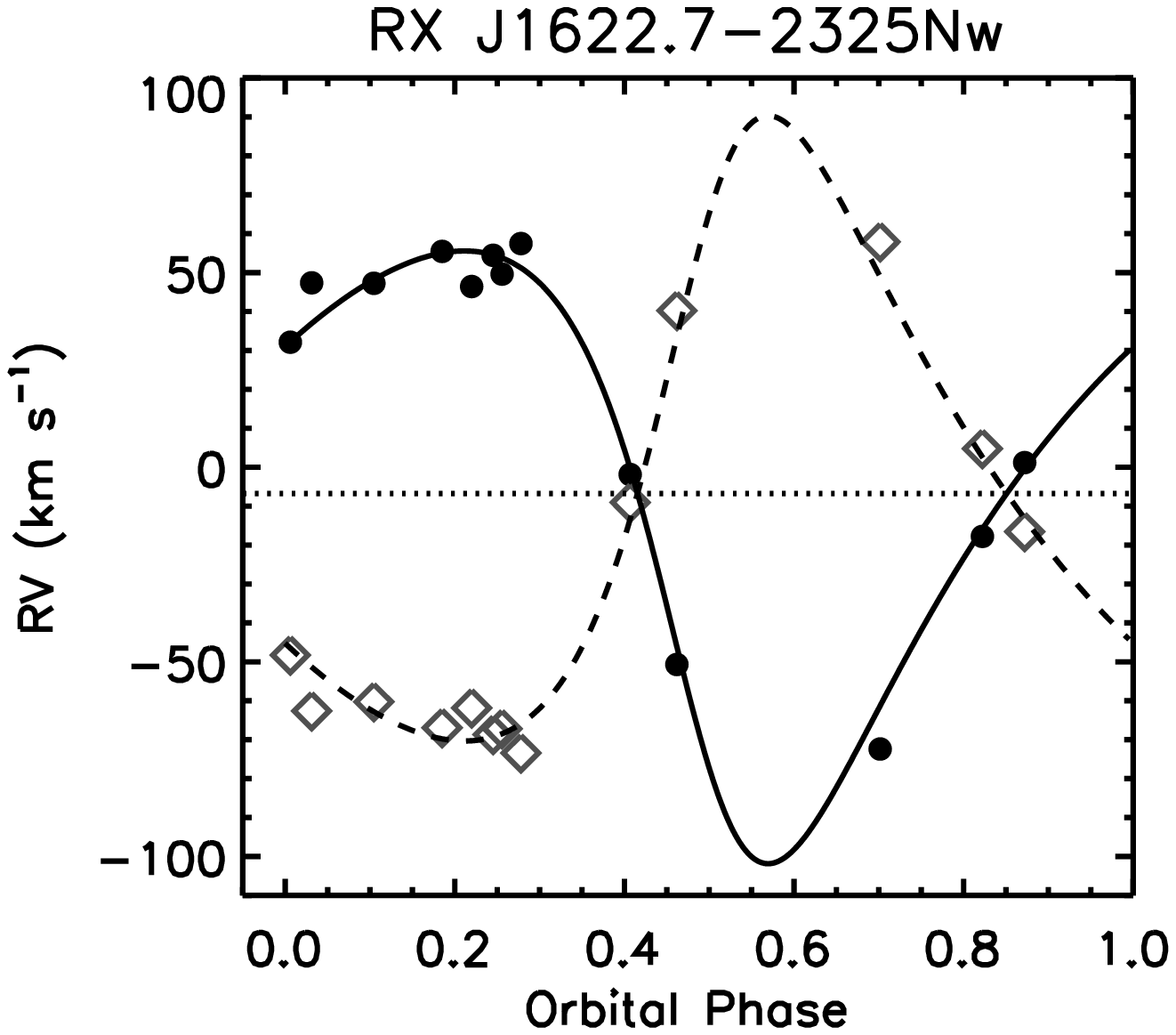}
\caption{Radial velocity as a function of phase for RX J1622.7-2325Nw. Measured radial velocities are represented as circles for the primary and as diamonds for the secondary. The dotted lined shows  the center of mass velocity, $\gamma$. Uncertainties in the RV are not shown because they are smaller than the symbol size.}\label{rxj_curve}
\end{figure}

Although  the primary versus secondary radial velocity plots  (shown in Figures  \ref{rox_wilson} and  \ref{rxj_wilson})  present high correlations (99\%  for both of the target systems) this does not ensure accurate orbital fits: the radial velocity versus phase solution for   ROXR1 14 shows a much better fit  than the solution for RX J1622.7-2325Nw (the $\chi^{2}$ of the orbital solutions are 1.1 and 21.2, respectively). This is seen  in Figures \ref{rox_curve} and \ref{rxj_curve}, where another parameter, the orbital phase, is also taken into account,  introducing an additional factor. Thus, the uncertainty in the orbital period also comes into play. It is unclear why our orbital fit for RX J1622.7-2325Nw is significantly worse than that for ROXR1 14. Some data points differ from the solution by more than $10 \sigma$. It is possible that another as-yet undetected low-mass star exists in the system, however, after subtracting the orbital solution shown in Figure \ref{rxj_curve}, no obvious periodic radial velocity modulation was apparent in the residual.

\begin{figure}[!h]
\centering
\includegraphics[scale=.60]{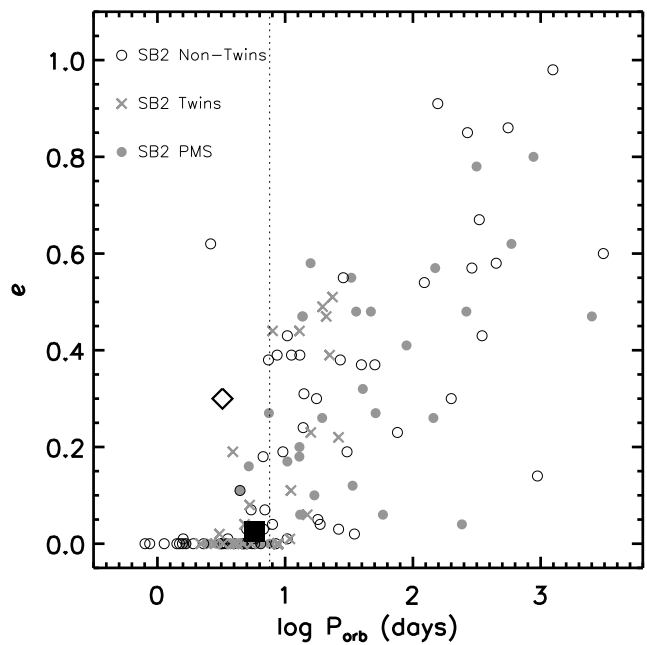}
\caption{Orbital eccentricity vs orbital period. Grey filled circles are PMS SBs from Melo et al. (2001). ROXR1 14 is the black filled square and RX J1622.7-2325Nw is the open diamond. Main sequence F and G star binaries from Simon \& Obbie (2009) are the grey crosses (twins of mass ratio unity) and the black open circles (non-twins).   The vertical, dotted line marks the circularization cut-off period (Zahn \& Boucher 1989).  The young and old populations are distributed throughout the same parameter space, implying the onset of circularization at a very young age.}
\label{ecce_fig}
\end{figure}

An interesting result of our analysis is the high eccentricity found for the short-period SB2 RX J1622.7-2325Nw.   The theory of tidal circularization introduced by  Zahn \& Bouchet (1989) predicts that low-mass stars have a cut-off circularization period, which for PMS stars is 7.56 days (Melo et al. 2001): systems with orbital periods less than 7.56 days are expected to have very low eccentricity. In Table 3 from Melo et al. (2001) it can be seen that in addition to RX J1622.7-2325Nw there are three PMS systems with orbital periods shorter than the cut-off circularization period with eccentric orbits (Figure \ref{ecce_fig}). However, SB2 RX J1622.7-2325Nw has an even shorter-period and higher eccentricity than these systems. Melo et al. (2001) comment that although currently there is no clear explanation for this behavior, there are some hypotheses that have been introduced in order to explain it. One is the  existence of low-mass companions, in higher-order multiple systems, giving rise to perturbations in the orbit. The other potential explanation is the presence of perturbing circumbinary disks.

Given the indistinguishable distribution of the MS spectroscopic binary twin and non-twin populations compared to the PMS population shown in Figure \ref{ecce_fig}, we do not believe that the existence of eccentric, short-period SBs is an age effect. Orbital circularization apparently at least begins for short-period systems early in a binary's  lifetime, likely in $< 1\,Myr$. All the eccentric, short-period systems in Figure \ref{ecce_fig} should be studied further for the possible presence of tertiary components and/or disks.\\

For our particular case,  we know already that RX J1622.7-2325Nw is one of the components of a hierarchical quadruple system (Prato 2007). The Ne component, itself a 0.1'' binary, is located $\sim$1$''$ away. At the $120\,pc$ distance to Ophiuchus (Loinard et al. 2008), this is equivalent to a projected separation of $108$\,AU. Thus one obvious mechanism for the high eccentricity in the RX J1622.7-2325Nw SB2 is the presence of this companion pair. Perturbation of the SB2 via the Kozai mechanism is possible if the outer orbit is highly inclined with respect to the SB2 orbit (Eggleton \& Kiseleva-Eggleton  2001;  Fabrycky \& Tremaine  2007). Although we cannot prove that this is the case, given the minimum period
of $>$1000 years implied by the separation, the Kozai effect is a well known phenomenon and provides a likely explanation.

Cieza et al. (2007) studied a sample of $\sim 230$ WTTS located in nearby star forming regions with the Spitzer Space Telescope and observed both RX J1622.7-2325Nw and ROXR1 14. Using the  IRAC (3.6 $\mu m$, 4.5 $\mu m$, 5.8 $\mu m$, and 8.0 $\mu m$ bands) and MIPS (24 $\mu m$) instruments they   searched for  IR excesses around  WTTS as an indicator of cool circumstellar dust disks in these systems. Results presented in Table 1 of Cieza et al. (2007) show no evidence for  cool, dusty disks around our SB2 systems. Therefore, the presence of the Ne component at $120$ AU away is probably the best explanation for the high eccentricity in the orbit of RX J1622.7-2325Nw. The perturbing  action of a closer multiple component, as yet undetected, is also possible.

\section{SUMMARY}\label{sum}
We obtained high-resolution IR spectra, with the NIRSPEC and Phoenix spectrometers, of the PMS SB2s ROXR1 14 and RX J1622.7-2325Nw, located in Ophiuchus.  Accurate orbital elements were calculated, using radial velocities derived from two-dimensional cross-correlation analysis, for both targets. Mass ratios of $q=0.97\pm 0.01$ and $q=0.98\pm 0.06$ and periods of $5.72$ days and $3.23$ days were determined for ROXR1 14 and RX J1622.7-2325Nw, respectively. The  eccentricity of $0.30\pm 0.04$ and a period shorter than  the cut-off for circularization make RX J1622.7-2325Nw an interesting system: the highest eccentricity short-period PMS SB known to date. The most reliable explanation of its eccentric orbit may be potential perturbations caused by the 0.1'' visual binary companion $\sim$1.0$''$ away, as  RX J1622.7-2325Nw is a member of a hierarchical quadruple system. We have increased by 4\%  the sample of PMS SBs with known orbital elements,  an important step on the way to refining the calibration of the current young star  theoretical models.\\

\acknowledgments
V. R. thanks  Lowell Observatory for the curricular practical training internship and the supportive environment provided.  We thank H. Roe for obtaining the spectra taken in July, 2009, and S. Porter for helpful discussions about the Kozai mechanism. We are grateful to the anonymous referee for a prompt and helpful report which improved the manuscript. Final stages of this research were supported by NSF grant AST-1009136 (to L.P.). This work made use of the SIMBAD reference database, the NASA Astrophysics Data System, and the data products from the Two Micron All Sky Survey, which is a joint project of the University of Massachusetts and the Infrared Processing and Analysis Center/California Institute of Technology, funded by the National Aeronautics and Space Administration and the National Science Foundation. Data presented herein were obtained at the W. M. Keck Observatory from telescope time allocated to the National Aeronautics and Space Administration through the agency's scientific partnership with the California Institute of Technology and the University of California. The Observatory was made possible by the generous financial support of the W.M. Keck Foundation. We recognize and acknowledge the significant cultural role that the summit of Mauna Kea plays within the indigenous Hawaiian community.  We are very  grateful for the opportunity to conduct observations from this special mountain. The work done in this paper is also based on observations obtained in
queue mode at the Gemini Observatory, and we appreciate the effort of Gemini science staff that made that possible.  Gemini Observatory is operated by the Association of Universities for Research in Astronomy, Inc., under a cooperative agreement with the NSF on behalf of the Gemini partnership: the NSF (US), STFC (UK), the NRC (Canada), CONICYT (Chile), ARC (Australia), MCT (Brazil) and MinCyT  (Argentina).


\clearpage


\clearpage


\clearpage


\clearpage


\clearpage


\clearpage


\clearpage


\clearpage



\clearpage
\begin{deluxetable}{l c c  }
\tabletypesize{\scriptsize}
\tablecaption{Target properties\label{target_prop}}
\tablewidth{0pt}
\tablehead{
			& ROXR1 14 & RX J1622.7-2325Nw \\
			}
\startdata
R.A (J2000.0)	& 16 26 03.3  &16 22 46.8  \\
Dec (J2000.0)	&  -24 17 47  &-23 25 33   \\
Spectral Type   &  M1  & M1\\
J\,\,\,(mag)	     		& $10.67\pm 0.02$    & $9.81\pm 0.03$\tablenotemark{a}  \\
H\,\,(mag)	  		& $9.58\pm 0.02$  &  $8.65\pm 0.03$\tablenotemark{a}   \\
K\,\,(mag)	 		&  $9.11\pm 0.02$ & $9.03\pm 0.06$ \\
$EW_{H\alpha}$ (\AA)\tablenotemark{b} 		&-2.1  & -1.7\\
$EW_{Li I}$ \,(\AA)\tablenotemark{b} 		&$0.64\pm 0.02$  &  $0.65\pm 0.05$\\
${\it v}$sin\emph i ($km\,s^{-1}$)   &$15\pm 3$ & $25\pm 5$\\
\enddata
\tablenotetext{a}{Includes light from close (0.1'') binary $\sim$1'' away. i.e. RXJ1622.7-2325Ne. K band AO images allowed us to apportion the light between the Ne and Nw components. therefore the K magnitude represents just the SB2.}
\tablenotetext{b}{ Mart\'in et al. 1998.}
\tablecomments{Units of right ascension are hours, minutes, and seconds, and units of declination are degrees, arcminutes, and arcseconds.}
\end{deluxetable}


\clearpage
\begin{deluxetable}{l r r r r r c  }
\tabletypesize{\scriptsize}
\tablecaption{ROXR1 14 Observations and Measured Velocities\label{roxdata}}
\tablewidth{0pt}
\tablehead{
\,\,\,\,\,\,\,\,\,\,\,\,\,\,\,\,\,\,JD & \multicolumn{1}{c}{$v_{1}$}  & \multicolumn{1}{c}{$\sigma_{v_{1}}$}  &  \multicolumn{1}{c}{$v_{2}$} &  \multicolumn{1}{c}{$\sigma_{v_{2}}$}  &\multicolumn{1}{c}{Phase} & UT Date  and \\                
\,\,\,\,(2,450,000+)    &   \multicolumn{1}{c}{$(km\,s^{-1})$}   &  \multicolumn{1}{c}{$(km\,s^{-1})$}   &  \multicolumn{1}{c}{$(km\,s^{-1})$}  &  \multicolumn{1}{c}{$(km\,s^{-1})$}   & & Instrument  \\
}
\startdata
2363.16 ......... &    0.47 & 1.125 & -17.59  & 1.094&0.2783& 2002 Mar 29 \tablenotemark{a}\\
2449.90 ......... &   32.51 & 1.031 & -50.31  & 1.027&0.4390&2002 Jun 24 \tablenotemark{a}  \\
2773.89 ......... &  -46.33 & 1.066 &  30.07  & 1.064&0.0673&2003 May 14 \tablenotemark{a}  \\
3124.97 ......... &   30.87 & 1.053 & -49.16  & 1.046&0.4305&2004 Apr 29 \tablenotemark{a}  \\
3424.18 ......... &   -8.27 & 1.268 &  -8.34  & 1.259&0.7276&2005 Feb 22 \tablenotemark{a}  \\
4220.96 ......... &  -50.89 & 1.027 &  35.97  & 1.015&0.9919&2007 Apr 30 \tablenotemark{a}  \\
4321.78 ......... &   22.99 & 1.089 & -39.34  & 1.056&0.6137&2007 Aug 09 \tablenotemark{a}  \\
4546.81 ......... &  -47.52 & 1.015 &  34.28  & 1.064&0.9453&2008 Mar 21 \tablenotemark{b}   \\
4599.66 ......... &  -23.58 & 1.103 &   7.07  & 1.142&0.1826&2008 May 13 \tablenotemark{b}     \\
4601.78 ......... &   32.08 & 1.041 & -48.68  & 1.033&0.5532&2008 May 15 \tablenotemark{b}     \\
4609.55 ......... &  -45.18 & 1.030 &  29.89  & 1.119&0.9112&2008 May 23 \tablenotemark{b}     \\
4630.55 ......... &   28.36 & 1.083 & -44.98  & 1.019&0.5817&2008 Jun 13 \tablenotemark{b}   \\
4632.75 ......... &  -49.54 & 1.017 &  35.41  & 1.036&0.9662&2008 Jun 15 \tablenotemark{b}   \\
4643.58 ......... &  -35.47 & 1.087 &  22.46  & 1.055&0.8591&2008 Jun 26 \tablenotemark{b}    \\
4646.57 ......... &   26.15 & 1.100 & -41.60  & 1.062&0.3817&2008 Jun 29 \tablenotemark{b}  \\
5039.84 ......... &  -38.08 & 1.060 &  22.14  & 1.070&0.1190&2009 Jul 27 \tablenotemark{a}        \\ 
\enddata

\tablenotetext{a}{NIRSPEC}
\tablenotetext{b}{Phoenix}
\end{deluxetable}


\clearpage
\begin{deluxetable}{l r r r r r r c  }
\tabletypesize{\scriptsize}
\tablecaption{RX J1622.7-2325Nw Observations and Measured Velocities\label{rxjdata}}
\tablewidth{0pt}
\tablehead{
\,\,\,\,\,\,\,\,\,\,\,\,\,\,\,\,\,\,JD &  \multicolumn{1}{c}{$v_{1}$}  & \multicolumn{1}{c}{$\sigma_{v_{1}}$} & \multicolumn{1}{c}{$v_{2}$}  & \multicolumn{1}{c}{$\sigma_{v_{2}}$} &\multicolumn{1}{c}{Phase} &  UT Date and\\                
\,\,\,\,(2,450,000+)     &  \multicolumn{1}{c}{ $(km\,s^{-1})$}	 &  \multicolumn{1}{c}{ $(km\,s^{-1})$}  &  \multicolumn{1}{c}{ $(km\,s^{-1})$} &   \multicolumn{1}{c}{$(km\,s^{-1})$} &  &Instrument    \\
}
\startdata
2449.80 ......... &  -1.87   & 1.438   &  -9.05 & 1.557&0.4095 &  2002 Jun 24 \tablenotemark{a}\\
2771.98 ......... &   47.37  & 1.234   & -62.62 & 1.098&0.0341 & 2003 May 12 \tablenotemark{a}\\
2804.93 ......... &   46.42  & 1.254   & -61.84 & 1.221&0.2227 & 2003 Jun 14 \tablenotemark{a}\\ 
3198.79 ......... &   32.11  & 1.421   & -48.26 & 1.377&0.0089 & 2004 Jul 12 \tablenotemark{a}\\
3424.18 ......... &  -72.40  & 1.411   &  57.89 & 1.265&0.6897 & 2005 Feb 22 \tablenotemark{a}\\
4321.79 ......... &   49.61  & 1.038   & -67.20 & 1.062&0.2585 & 2007 Aug 09 \tablenotemark{a}\\
4599.69 ......... &   55.44  & 1.065   & -66.94 & 1.167&0.1879 & 2008 May 13  \tablenotemark{b}\\
4600.58 ......... &  -50.68  & 1.267   &  40.23 & 1.277&0.4647 & 2008 May 14 \tablenotemark{b} \\
4601.75 ......... &  -17.78  & 1.733   &   4.76 & 1.464&0.8249 & 2008 May 15  \tablenotemark{b}\\
4602.66 ......... &   47.25  & 1.298   & -60.28 & 1.418&0.1074 & 2008 May 16 \tablenotemark{b} \\
4609.58 ......... &   54.44  & 1.176   & -68.68 & 1.310&0.2480 & 2008 May 23 \tablenotemark{b}  \\
4611.61 ......... &   1.18   & 1.703   & -16.55 & 1.518&0.8748 & 2008 May 25 \tablenotemark{b} \\
5039.81 ......... &   57.46  & 1.142   & -73.42 & 1.122&0.2807 & 2009 Jul 27 \tablenotemark{a} \\
\enddata                                
\tablenotetext{a}{NIRSPEC}
\tablenotetext{b}{Phoenix}
\end{deluxetable}


\clearpage
\begin{deluxetable}{l r r c c  }
\tabletypesize{\scriptsize}
\tablecaption{Standard Stars\label{standard_st}}
\tablewidth{0pt}
\tablehead{
\multicolumn{1}{c}{Object} & \multicolumn{1}{c}{RA} & \multicolumn{1}{c}{Dec} & \multicolumn{1}{c}{RV\tablenotemark{a}} & \multicolumn{1}{c}{Spectral} \\
&  \multicolumn{1}{c}{(J2000.0)} & \multicolumn{1}{c}{(J2000.0)} & \multicolumn{1}{c}{($km$ $s^{-1}$)} & \multicolumn{1}{c}{Type\tablenotemark{a} }\\
}
\startdata

BS 8086     & 21 06 55.3 &  +38 44 31  & -65.0 &  K7 \\
GL 763       & 19 34 39.8 & +04 34 57   & -60.5      & M0 \\
\enddata
\tablenotetext{a}{Prato et al. 2002b.}
\end{deluxetable}

\clearpage
\begin{deluxetable}{l c c }
\tabletypesize{\scriptsize}
\tablecaption{Orbital Elements and Related Parameters for RX J1622.7-2325Nw and 
ROXR1 14  \label{parameters}}
 \tablewidth{0pt}
\tablehead{
Element			& ROXR1 14 & RX J1622.7-2325Nw \\
}
\startdata
P \,\,\,\,(d)	& $5.72$\tablenotemark{a}   &$3.23$\tablenotemark{a}    \\
$\gamma$ \,\,\,\,(km s$^{-1}$) &  $-7.98\pm 0.18$  & $-6.75\pm 1.09$  \\
$K_{1}$ (km s$^{-1}$)      &  $42.66\pm 0.33$  & $78.71\pm 3.52$\\
$K_{2}$ (km s$^{-1}$)      &  $43.94\pm 0.33$  & $80.31\pm 3.58$\\
{\it  e}   		& $0.020\pm 0.007$ & $0.30\pm 0.037$\\
$\omega$	 (deg) 	& $3.87\pm 17.04$   & $133.45\pm 4.41$  \\
T (JD)		& $2452175.62\pm 0.27$ & $2452019.97\pm 0.03$\\
$q=M_{2}/M_{1}$  & $0.97\pm 0.01$      &     $0.98\pm 0.06$\\
$a_{1}sin$\emph i (gm)	&$3.36\pm 0.03$ &  $3.34\pm 0.14$\\
$a_{2}sin$\emph i (gm)    &$3.46\pm 0.03$  &  $3.40\pm 0.15$\\
\# of  Observations        &    16 &  13 \\
\enddata

\tablenotetext{a} {The observations span hundreds of orbits and therefore periods are very accurately determined.  The formal errors from the least squares fitting are only a few seconds, which may be an underestimate.  However, it is likely that the error is quite small.}
\end{deluxetable}



\begin{thebibliography}{}

\bibitem[bean06]{bean06} Bean, J. L., Sneden C., Hauschildt P. H., Johns-Krull C. M., \& Benedict G. F. 2006, \apj, 652, 1604

\bibitem[Bender \& Simon(2008)]{Bender08} Bender, C.~F., \& Simon, M.\ 2008, \apj, 689, 416

\bibitem[casanova95]{Casanova95} Casanova, S., Montmerle, T., Feigelson, E. D., Andr\'e, P. 1995, \apj, 439, 752

\bibitem[Cieza07]{Cieza07} Cieza, L., et al. 2007, \apj, 667, 308

\bibitem[Eggleton01]{Eggleton01} Eggleton, P. P., \& Kiseleva-Eggleton, L.  2001, \apj, 562, 1012

\bibitem[Fabrycky07]{Fabrycky07} Fabrycky, D., \& Tremaine, S. 2007, \apj, 669, 1298

\bibitem[hillen]{Hillen04} Hillenbrand, L. A., \& White, R. J. 2004, \apj, 604, 741

\bibitem[Hin03]{Hin03} Hinkle, K. H., et al. 2003, Proc. SPIE, 4834, 353

\bibitem[Kurtz99]{Kurtz99} Kurtz, M. J., Mink, D. J., Wyatt, W. F., Fabricant, D. G., Torres, G., Kriss, G. A., \& Tonry, J. L. 1992, in Astronomical Data Analysis Software and Systems I, ed. D.M. Worral, C. Biemesderfer \& J. Barnes (ASP Conf. Ser., 25), 432

\bibitem[Loinard08]{Loinard08} Loinard, L., Torres, R., Mioduszewski, A., \& Rodr\'iguez, L. 2008, \apj, 675, L29

\bibitem[Martin98]{Martin98} Mart\'in, E. L., Montmerle, T., Gregorio-Hetem J., \& Casanova, S., 1998, \mnras, 300, 733 

\bibitem[Mazeh90]{Mazeh90} Mazeh, T. 1990, \aj, 99, 675

\bibitem[McLean00]{McLean00} McLean , I. S, Graham, J. R., Becklin, E. E. Figer, D. F., Larkin, J. E., Levenson, N. A., \& Teplitz, H. I. 2000, Proc. SPIE, 4008, 1048

\bibitem[McLean08 ]{McLean08} McLean , I. S., et al. 1998, Proc. SPIE, 3354, 566

\bibitem[Melo01]{Melo01} Melo, C. H. F., Covino, E., Alcal\'a, J. M., \& Torres, G. 2001, \aap, 378, 898

\bibitem[Prato02a]{Prato02a} Prato, L., Simon, M., Mazeh, T., Zucker S., \& McLean I. S. 2002a, \apj, 579, L99

\bibitem[Prato02b]{Prato02b} Prato, L., Simon, M., Mazeh, T., McLean I. S., Norman, D., \& Zucker S. 2002b, \apj, 569, 863

\bibitem[Prato07]{Prato07} Prato, L.  2007, \apj, 657, 338

\bibitem[Press92]{Press92} Press, W. H., Teukolsky, S. A., Vetterling, W. T., \& Flannery, B. P. 1992, Numerical Recipes in Fortran: The Art of Scientific Computing (2nd ed,; Cambridge: Cambridge Univ. Press)

\bibitem[Rousselot et al 2000]{Rousselot00} Rousselot, P., Lidman, C., Cuby, J.-G., Moreels, G., \& Monnet, G. 2000, \aap, 354, 1134

\bibitem[Shaefer08]{Schaefer08} Schaefer, G. H., Simon, M., Prato, L., \& Barman, T.  2008, \aj, 135, 1659

\bibitem[Simon04]{Simon04} Simon, M., \& Prato, L. 2004, \apj, 613, L69

\bibitem[Simon09]{Simon09} Simon, M., \& Obbie R. C. 2009, \aj, 137, 3442

\bibitem[Steffen01]{Steffen01} Steffen, A., et al. 2001, \aj, 122, 997

\bibitem[Thorstensen01]{Thorstensen01} Thorstensen,  J.R. 2001, Skycalc UserÕs Manual, (Tucson: NOAO, contributed software)

\bibitem[Wilson41]{Wilson41} Wilson, O. C. 1941, \apj, 93, 29


\bibitem[Zahn89]{Zahn89} Zahn, J.-P., \&  Bouchet, L., 1989, \aap, 223, 112  

\bibitem[Zucker94]{Zucker94} Zucker, S., \& Mazeh, T. 1994, \apj, 420, 806

\end{thebibliography}
\end{document}